\documentstyle[aps,twocolumn,epsf,amssymb,prl]{revtex}
\begin{document}
\bibliographystyle{prsty} \title{Suppression of $\bbox{T_c}$ in
  superconducting amorphous wires} \author{Yuval Oreg$^a$ and
  Alexander M. Finkel'stein$^b$ \protect{\cite{Lan}}} \address{$^a$
  Lyman Laboratory of Physics, Harvard University,
  Cambridge MA 02138, USA\\
  $^b$Department of Condensed Matter Physics The Weizmann\\
  Institute of Science 76100 Rehovot, Israel \\
  {}~{ }
  \medskip \\
  \parbox{14cm} {\rm The suppression of the mean field temperature of
    the superconducting transition, $T_c$, in homogeneous amorphous
    wires is studied.  We develop a theory that gives $T_c$ in
    situations when the dynamically enhanced Coulomb repulsion
    competes with the contact attraction. The theory accurately
    describes recent experiments on $T_c$--suppression in
    superconducting wires, after a procedure that minimizes the role
    of nonuniversal mechanisms influencing $T_c$ is
    applied. \smallskip} \vspace{-0.3cm}}
\maketitle
Disorder suppresses the superconductivity transition in
morphologically homogeneous superconductors~\cite
{SC2D:Maekawa82,SC2D:Takagi82,SC2D:Finkelstein8788,DSC:Smith95}
because the diffusive character of the electron motion in dirty
systems makes the Coulomb interaction more
effective~\cite{DS:AltLee80}. As a result, the attraction between the
electrons in Cooper pairs becomes weaker, and the transition
temperature, $T_{c}$, is lowered. In two dimensions ($2D$) the
influence of disorder on $T_{c}$ can be studied systematically by
varying the film thickness $d$
\cite{SC2D:Graybeal84,SC2D:Haviland89,SC2D:Valles89}.  In uniform
films $T_{c}$, being well defined, is suppressed as the sheet
resistance, $R_{\Box }$, increases with decreasing $d$. (For a review
see Ref.~\cite{SC2D:Finkelstein94}.) When the geometry of the sample
is such that its dimension is lowered towards the one-dimensional
($1D$) limit, the suppression of superconductivity should become more
pronounced~\cite{SCW:Ebisawa86}.

Recently, efforts have been made~\cite
{SCW:Graybeal87,SCW:Sharifi93,SCW:Xiong97} to extend the experiments
in films to narrow wires by fabricating a series of amorphous $Pb$
wires of different thicknesses and widths. It has been found that the
$T_{c}$-suppression becomes stronger as the wires' width reduces below
$1000$\AA.  The experiment of Refs.~\cite{SCW:Sharifi93,SCW:Xiong97}
is in the crossover region from $2D$ to $1D$. Actually, the wires are
in the $1D$ limit as far as superconducting fluctuations are concerned
\cite {DR:Aslamazov68}, but they are in the crossover region from $2D$
to $1D$ with respect to the diffusive motion of the electrons.

From the theoretical point of view, the problem of $T_{c}$-suppression
in $1D$ wires is rather intriguing. As is well known, the
superconductivity transition is determined by a series of
logarithmically divergent terms describing the electron scattering in
the two-particle Cooper channel. In $2D$ systems the corrections due
to the electron-electron ($e$--$e$) interactions combined with
disorder are logarithmically divergent as well \cite{DS:AltLee80}. As
the whole problem is controlled by logarithmic singularities, it can
be studied by renormalization group (RG)
methods~\cite{SC2D:Finkelstein84a}.  In $1D$, due to the reduced
dimensionality, the effect of $e$--$e$ interactions is more singular.
It produces corrections that diverge as the square root of the
frequency. The presence of two types of singularities demands a
special analysis in the calculation of $T_{c}$.  In this paper we
develop a theory that describes adequately the effect of the
dynamically enhanced $e$--$e$ interaction on $T_{c}$ in the crossover
region from $2D$ to $1D$ and perform a detailed comparison with the
experiment.

\begin{figure}[t]
  \vglue 0cm \hspace{0.05\hsize} \epsfxsize=0.8\hsize
  \epsffile{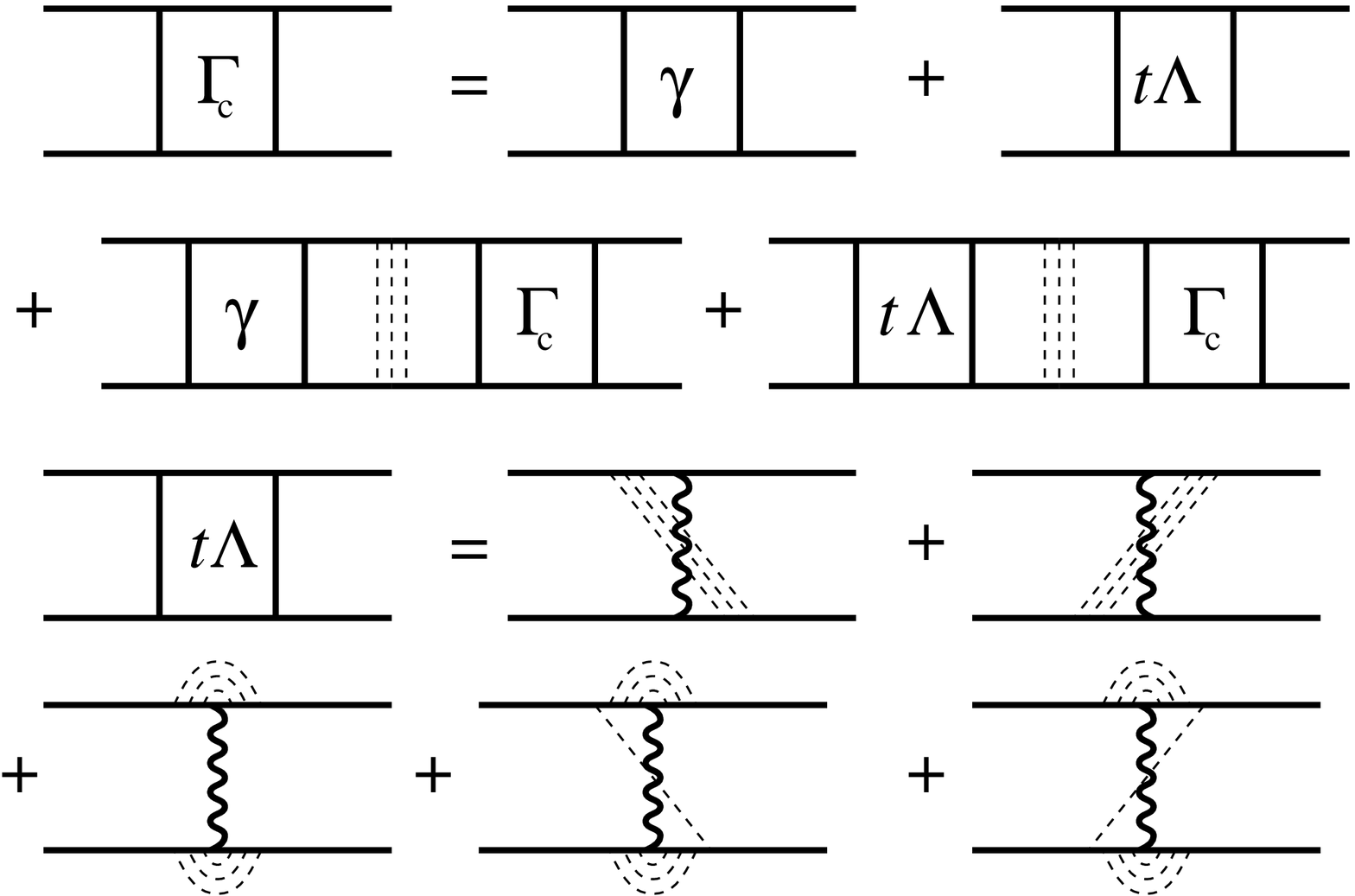}
  \refstepcounter{figure} \label{fg:CC} \\
  {\small FIG.\ \ref{fg:CC} The diagrammatic equation for the
  scattering amplitude $\Gamma_c$ in the Cooper channel.  The block
  $\gamma $ denotes the BCS-interaction amplitude. The block $t\Lambda
  $ describes the interplay of the Coulomb interaction with disorder
  that leads to the suppression of $T_c$.  The wavy line is the
  screened Coulomb interaction, dashed lines describe impurity
  scattering.}
\end{figure}

The mean field temperature, $T_{c}$, is defined as the temperature at
which the electron scattering amplitude in the Cooper channel,
$\Gamma_{c}$, becomes infinite. Fluctuations of the superconductivity
order parameter lead to a broadening of the phase transition. However,
its mean field temperature can be found experimentally by fitting the
upper part of the resistive transition to the Aslamazov--Larkin theory
\cite {DR:Aslamazov68}.  The diagrammatic representation of the
amplitude $\Gamma _{c}$ is shown in Fig~\ref{fg:CC}. In addition to
the contact BCS-interaction amplitude $\gamma$, the terms arising as a
result of the interplay of the Coulomb interaction and disorder are
also included in the Cooper ladder-diagram series.  [The impurity
scattering does not influence the $e$--$e$ interaction mediated by
phonons because in the long wavelength limit the lattice defects
oscillate together with the ions \cite{SC:Schmid73}.]  The resulting
equation for $\Gamma _{c}$ is:
\begin{eqnarray}
\Gamma _{c}(\epsilon _{n},\epsilon _{l})=-|\gamma |+t\Lambda (\epsilon
_{n}+\epsilon _{l}) \quad \quad \quad  \nonumber \\
-2\pi T\sum_{m=0}^{M}\left[ -|\gamma |+t\Lambda (\epsilon
_{n}+\epsilon _{m})\right] \nonumber \\
\times \frac{1}{\epsilon _{m}}\Gamma _{c}(\epsilon
_{m},\epsilon _{l}),  \label{eq:gamma}
\end{eqnarray}
where $\epsilon_{m}=2\pi T(m+1/2)$ is the Matsubara frequency, and the
summation over $m$ is limited by $M=\left( 2\pi T\tau \right)
^{-1}$. In this equation $\gamma$, the bare value of the amplitude
$\Gamma _{c}$, is rescaled in such a way that the Debye frequency as a
cut off energy is substituted by $\tau^{-1}$, the inverse of the
scattering time.  Then, $\gamma=1/ \ln\left(T_{c0}\tau /1.14\right),$
where $T_{c0}$ is the temperature of the superconducting transition in
the bulk limit. The parameter $t=(e^{2}/2\pi^{2}\hbar)R_\Box$
characterizes the level of disorder in a sample, where $R_\Box$ is the
sheet resistance.  The amplitude $\Lambda $ describing the combined
action of the $e$--$e$ interaction and disorder is given by
\begin{equation}
\Lambda (\omega _{n})=u\frac{4\pi D}{La}\sum_{q_{L}, q_{a}}\frac{1}{
Dq_{L}^{2}+Dq_{a}^{2}+\omega _{n}},  \label{eq:f}
\end{equation}
where $a$ and $L$ are the width and the length of the wire,
respectively.  The parameter $u$ describes the amplitude of the
$e$--$e$ interaction when the momentum $q$ transferred by this
interaction is not too small compared with the transferred frequency
$\omega $, namely when $q \gtrsim q_{\omega } = \sqrt{\omega /D}$. (As
was explained in Refs.~\cite
{SC2D:Finkelstein8788,SC2D:Finkelstein94}, the most divergent
contributions from the region $q<$ $ q_{\omega }$ cancel each other
out.  In this region of small momenta, the $e$--$e$ interaction
depends only on the frequency, and therefore it can be gauged out.)
Next, for amorphous $Pb$ films the spin-orbit scattering time is
expected to be only a few times longer than the elastic scattering
time and therefore the part of the $e$--$e$ interaction related to
spin density fluctuations can be neglected.  In that case, we may take
$u$ to be the value of the screened Coulomb interaction amplitude in
the region of momenta $q\gtrsim q_{\omega }$, which gives $u\approx
1/2$.

In $2D$ the summation in Eq.~(\ref{eq:f}) yields $\Lambda ({\omega }
_{n})\cong u\ln ({1/\omega }_{n}{\tau).}$ Therefore,
Eq.~(\ref{eq:gamma}) combines the usual BCS logarithms together with
the ones arising due to disorder. Unlike the ladder diagrams in the
BCS-theory, the integrations in the different blocks of the diagrams
in Fig.~\ref{fg:CC} cannot be factorized, because $\Lambda \left(
  \epsilon _{n}+\epsilon _{m}\right) $ matches the frequency arguments
of two neighboring blocks.  In order to solve this parquet-like
equation with a logarithmic accuracy one uses the approximation $\ln
(z+z^{\prime }) \cong \ln (\max \{z,z^{\prime }\})$, see e.g.
Ref.~\cite{XR:Nozieres69a}. Then, it is possible to apply the
``maximum section'' method. This procedure leads to the RG equation
for the amplitude
$\Gamma_{c}(\varepsilon,\varepsilon)$~\cite{SC2D:Finkelstein8788,SC2D:Finkelstein84a}:
$ d\Gamma _{c} / dl_{\varepsilon }= ut - \Gamma_{c}^{2},$ where
$l_{\varepsilon }=\ln (1/\varepsilon \tau )$. The integration of the
RG equation gives the suppression of $T_{c}$ by the Coulomb
interaction in $2D$ disordered systems:
\begin{equation}
\ln \left( \frac{T_{c}}{T_{c0}}\right) =\frac{1}{|\gamma|
}-\frac{1}{2 \sqrt{ ut }} \ln { \frac{ 1+ \sqrt{ ut }/\left| \gamma
\right| } { 1-\sqrt{ut} /\left| \gamma \right| }}.
\label{eq:2DTc}
\end{equation}
This formula accurately describes the experimental results in $MoGe$
films \cite{SC2D:Graybeal84}, with $u=1/2$ and using only one fitting
parameter, $\gamma$ \cite{SC2D:Finkelstein8788}.

In $1D$ the result of the summation in Eq.~(\ref{eq:f}) yields a
square root singularity in the amplitude $\Lambda(\omega_n) $. When
one deals with singularities stronger than logarithmic ones, the
approximations of the maximum section method cease to be valid, and a
different method should be invented.  In this Letter we treat the
problem of finding $T_{c}$ from Eq.~(\ref{eq:gamma}) as a sort of an
eigenvalue problem, which leads to an implicit equation for
$T_{c}$. To see this, we will consider $\Gamma _{c} ( \epsilon_{n} ,
\epsilon _{m} )$ as the matrix elements of a matrix
$\hat{\Gamma}_{c}$, and will write the solution of
Eq.~(\ref{eq:gamma}) for $\Gamma _{c}$ in matrix notations:
\begin{equation}
\hat{\Gamma}_{c}=\hat{\epsilon}^{\frac{1}{2}}\left( \hat{I}-|\gamma |\hat{\Pi
}\right) ^{-1}\hat{\epsilon}^{-\frac{1}{2}}\left( -|\gamma | \hat 1+t\hat{\Lambda}
\right) .  \label{eq:mat}
\end{equation}
Here $\hat{\Pi}(T)=\hat{\epsilon}^{-1/2} [ \hat{1}-|\gamma |^{-1} t
\hat{ \Lambda} ] \hat{\epsilon}^{-1/2}$, $ \hat {\epsilon}_{nm} =
\delta_{nm} (n+1/2)$, $\hat{\Lambda}_{nm}= \Lambda ( \epsilon_{n} +
\epsilon_{m})$, $ \hat{1}_{nm}=1,$ and $\hat{I}$ is a unit
matrix. Eq.~(\ref{eq:mat}) is written in such a form that $\hat{\Pi}$
is a symmetric matrix. Notice, that the dependence of $\hat{\Pi}$ on
the temperature $T$ is not only through the dependence of
$\hat{\Lambda}$ on the Matsubara frequencies, but also through the
matrix rank $M=\left( 2\pi T\tau \right) ^{-1}.$ The amplitude $\Gamma
_{c}$ diverges when the temperature is such that one of the
eigenvalues of the matrix $\hat{\Pi}(T)$ is equal to $ |\gamma
|^{-1}$, i.e., at $T=T_c$ the equation
\begin{equation}
\left[ |\gamma |^{-1}\hat{I}-\hat{\Pi}(T_{c})\right] \left| \Psi
\right\rangle =0  \label{eq:es}
\end{equation}
holds. Thus, the equation determining $T_{c}$ can be obtained from an
eigenvalue problem. [One can also obtain an equation for $T_c$ by
considering a BSC-like gap equation with frequency dependent
interaction vertex, $-|\gamma| + t \Lambda$.] The matrix elements of
$\hat{\Pi}=\hat{\Pi} ^{0}+\hat{\Pi}^{1}$ are
\begin{eqnarray}
\hat{\Pi}_{nm}^{0}=[(n+1/2)(m+1/2)]^{-1/2},\nonumber \;\;\;\;\;\;\;\;\;\;\;\; \\ \hat{\Pi}
 _{nm}^{1}=-t[(n+1/2)(m+1/2)]^{-1/2}|\gamma |^{-1}\Lambda
 (\epsilon _{n}+\epsilon _{m}).  \label{eq:me}
\end{eqnarray}
As the matrix elements $ \hat{\Pi}_{nm}^{0}$ are factorized with
respect to $n$ and $m$, all the eigenvalues of the matrix
$\hat{\Pi}^{0}$, except one, are degenerate and equal to zero. The
eigenvector corresponding to the nonzero eigenvalue is $\Psi
_{n}^{0}=c/ \sqrt{n+1/2}$, and the equation $|\gamma |^{-1}\Psi
_{n}^{0}=\sum_{m}\hat{\Pi }_{nm}^{0}\Psi _{m}^{0}$ leads to the BCS
relation for $T_{c0}$:
\begin{equation}
|\gamma |^{-1}=l_{0}(T_{c0}), \quad l_{0}(T)\equiv
\sum_{m=0}^{M}\frac{1}{ m+1/2}=\ln \frac{1.14}{T\tau }.
\label{eq:tco}
\end{equation}
Our strategy now will be to calculate the corrections to this
eigenvalue perturbatively in $\hat{\Pi}^{1}$ (notice that
$\hat{\Pi}^{1}\varpropto t$), and in this way to get an implicit
equation for $T_{c}$. Since $\hat{\Pi}$ is symmetric we can perform
this program using a standard perturbation theory:
\begin{equation}
|\gamma |^{-1}=l_{0}(T)+l_{1}(T)+l_{2}(T)+ \dots  \label{eq:l0l2}
\end{equation}
The first order term can be obtained straightforwardly
\begin{eqnarray}
l_{1}=\left\langle \Psi ^{0}\left| \hat{\Pi}^{1}\right| \Psi
^{0}\right\rangle \equiv -\frac{t}{l_{0}|\gamma |}\Sigma _{2}(T), \nonumber  \\
 \Sigma_{2}(T)=\sum_{n,m=0}^{M}\frac{\Lambda \left( \epsilon _{n}
+\epsilon_{m}\right) }{(n+1/2)(m+1/2)}.  \label{eq:l1}
\end{eqnarray}
The prefactor $1/l_{0}$ appears in $l_1$ because the normalization
factor $c$ of the eigenvector $\Psi _{n}^{0}$ is equal to
$1/\sqrt{l_{0}}$. Since all the eigenvalues of the operator
$\hat{\Pi}^{0}$ are degenerate except the one under studying, it is
also possible to find the higher order corrections using only the
eigenvector $\left| \Psi ^{0}\right\rangle $, without involving other
eigenvectors.  We demonstrate it here for the second order term, but a
generalization to higher orders is straightforward.  In the second
order
\begin{eqnarray}
l_{2}=\sum_{i\neq 0}\frac{\left\langle \Psi ^{0}\left| \hat{\Pi}^{1}\right|
\Psi ^{i}\right\rangle \left\langle \Psi ^{i}\left| \hat{\Pi}^{1}\right|
\Psi ^{0}\right\rangle }{l_{0}} \nonumber \\
=\frac{1}{l_{0}}\left[ \left\langle \Psi
^{0}\left| \hat{\Pi}^{1}\hat{\Pi}^{1}\right| \Psi ^{0}\right\rangle -\left(
l_{1}\right) ^{2}\right] ,  \label{eq:l2}
\end{eqnarray}
where 
\begin{eqnarray}
\left\langle \Psi ^{0}\left| \hat{\Pi}^{1}\hat{\Pi}^{1}\right| \Psi
^{0}\right\rangle \equiv \frac{t^{2}}{l_{0}|\gamma |^{2}}\Sigma_{3}(T),\quad
\nonumber \\
\Sigma_{3}(T)=\sum_{nmk=0}^{M}\frac{\Lambda \left( \epsilon _{n}+\epsilon
_{k}\right) \Lambda \left( \epsilon _{k}+\epsilon _{m}\right) }{%
(n+1/2)(m+1/2)(k+1/2)}.  \label{eq:sigmatr}
\end{eqnarray}
Inverting Eq.~(\ref{eq:l0l2}) perturbatively in $t$ and having in mind that $
|\gamma |l_{0}(T_{c0})=1$, we find 
\begin{eqnarray}
\ln  \frac{T_{c}}{T_{c0}} =-t\Sigma_{2}(T_{c0}) \;\;\;\;\;\;\;\;\; \;\;\;\;\;\;\;\;\; \nonumber \\
+t^{2}\left( 
\Sigma_{3}(T_{c0})+T_{c0}\left. \frac{\partial \Sigma_{2}(T)}{\partial T}
\right|_{T=T_{c0}}\Sigma_{2}(T_{c0}) \right)+ \dots  \label{eq:Tc}
\end{eqnarray}
Since Eq.~(\ref{eq:Tc}) gives an approximation for $\ln
(T_{c}/T_{c0})$, while the measured quantity in experiments is
$T_{c}/T_{c0}$, the first two terms of the perturbative series are
sufficient for the description of the $T_c$ suppression, if the
parameter $t$ is not too close to a critical value where $T_c$
vanishes. [The parameter $t$ should be inside the radius of
convergence of the series~(\ref{eq:Tc}). Outside this radius the
superconductivity is completely suppressed.]  In the $2D$ case
Eq.~(\ref{eq:Tc}) reproduces the first two terms of the expansion of
the right hand side of Eq.~(\ref{eq:2DTc}) in powers of $ut/\gamma
^{2}$:
\begin{equation}
\ln \left( \frac{T_{c}}{T_{c0}}\right) =\sum_{n=1}^{\infty }\frac{1}{
(2n+1)\gamma }\left( \frac{ut}{\gamma ^{2}}\right)^{n}. \label{eq:exp}
\end{equation}
We note that expansion (\ref{eq:exp}) does not contain a term $\propto
t^2/\gamma^4$. There are several diagrams that gives contributions to
that order, however, finally they cancel each other
\cite{SCW:Aleiner98}.  The main advantage of Eq.~(\ref{eq:Tc}) is that
it is not restricted to a logarithmic accuracy, and can be applied to
the description of the crossover from $2D$ to $1D$ systems.

\begin{figure}[t]
  \vglue -1.3cm \hspace{0.05\hsize} \epsfxsize=1 \hsize
  \epsffile{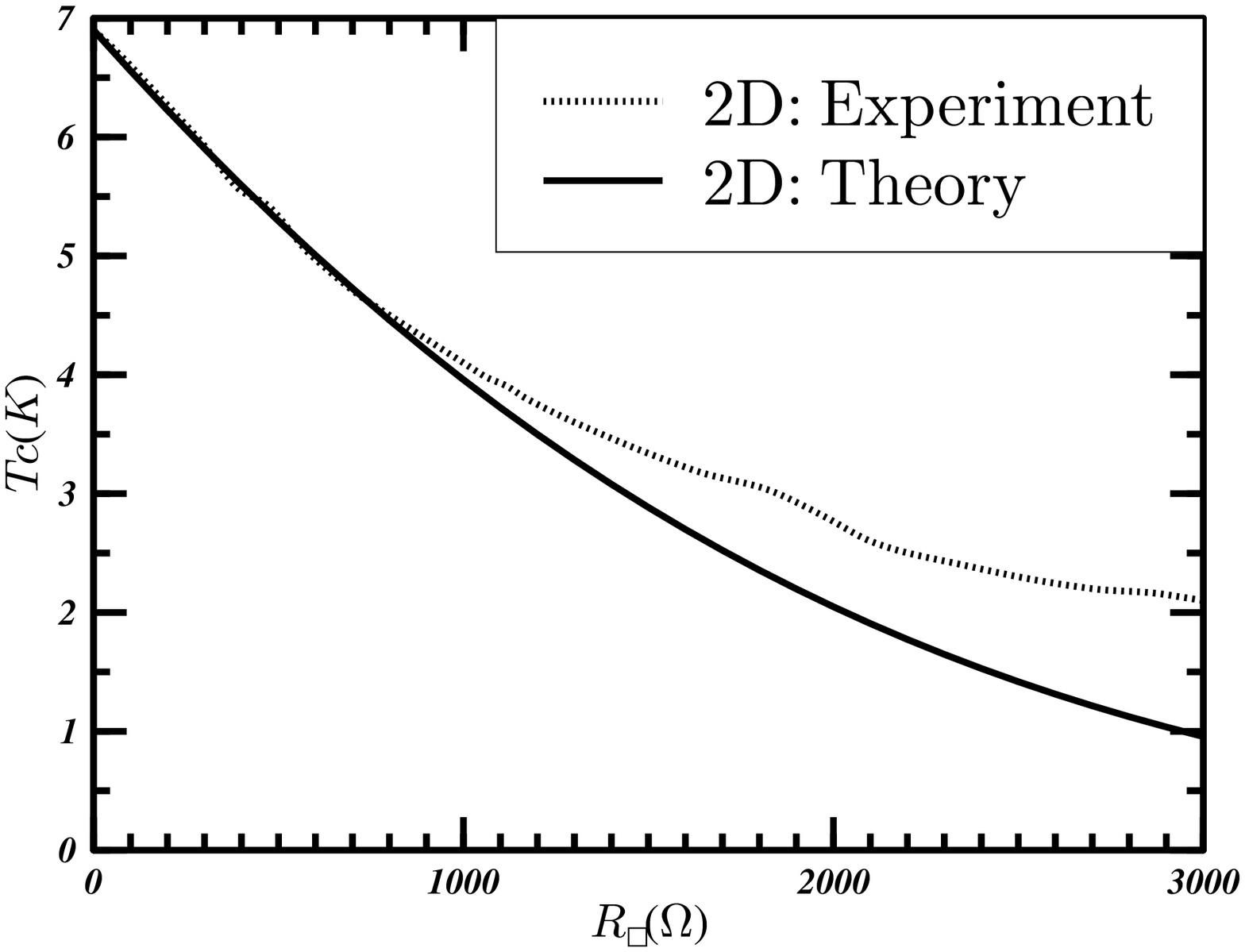} \refstepcounter{figure} \label{fg:et2D}\\
\vglue -1.3cm
  {\small FIG.\ \ref{fg:et2D} Comparison between the theory (solid
  line) and the experimental results \protect{\cite{SCW:Xiong97}}
  (dashed line) in $Pb$ films. The fitting parameter $\gamma =
  -0.16$ is determined from the initial slope of $T_{c}(R_{\Box })$.
  The deviation between the theory and the experiment at large
  $R_\Box$ shows that the interplay of the Coulomb interaction and
  disorder is not the only mechanism influencing $T_c$.}
\end{figure}

\begin{figure}[t]
\vglue 0cm
\hspace{0.01\hsize}
\epsfxsize=0.8 \hsize
\epsffile[80 250 455 720]{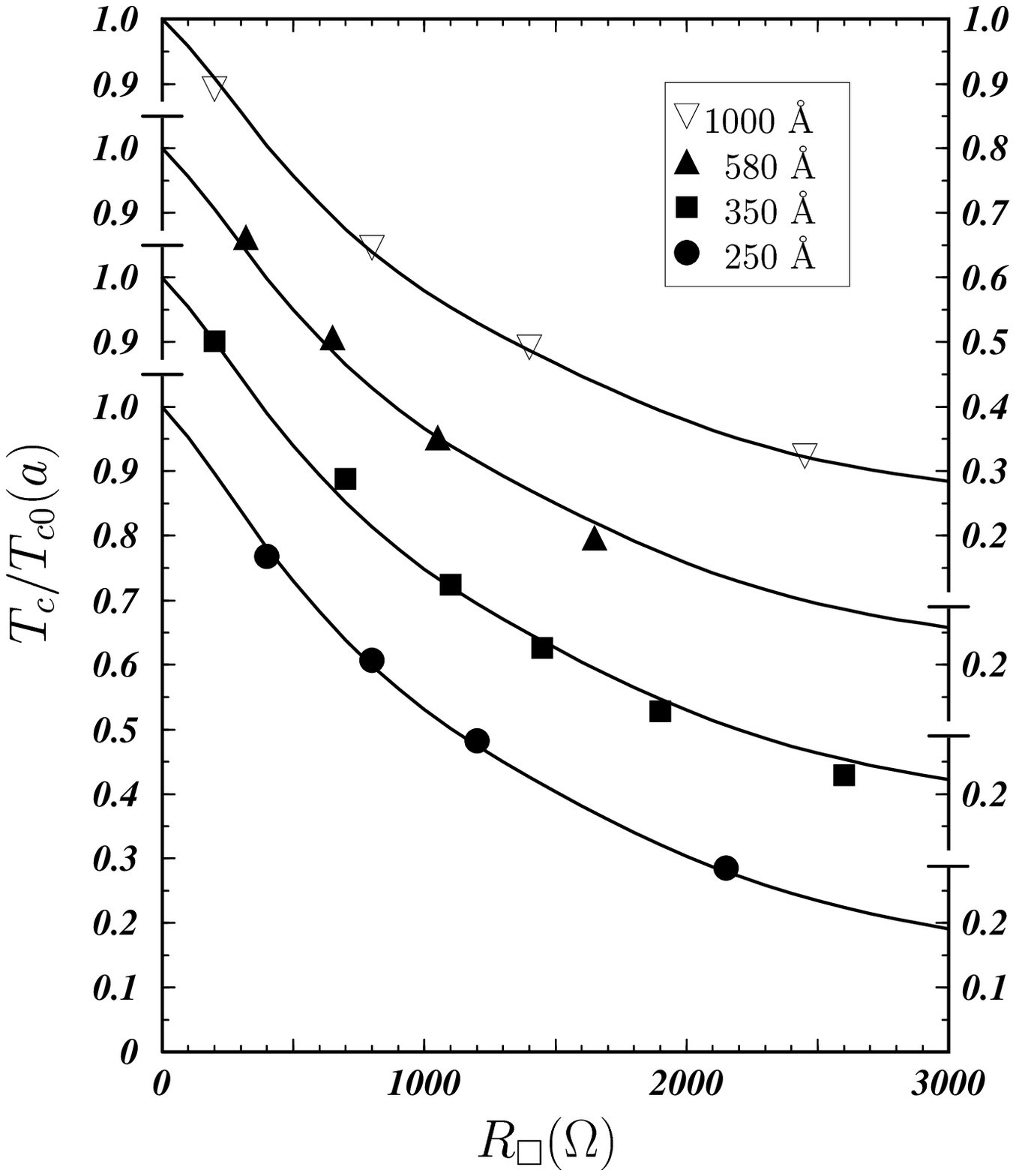}
\refstepcounter{figure} \label{fg:et250} \\
 {\small FIG.\ \ref{fg:et250} Comparison between the theory (solid
line) and the experimental data \protect{\cite{SCW:Xiong97}} for wires
of different width.  The short range effects are excluded assuming
that they are the same as in $2D$.  For each wire width, $a$, the
theoretical curves and the experimental data points are normalized by
the extrapolated value $T_{c0}(a) = T_c(R_\Box \rightarrow 0)$.  For
all width we use $\gamma =-0.16$ as in $2D$.}
\end{figure}

 In the experiment of Xiong et al.~\cite{SCW:Xiong97} the mean field
temperature of the superconducting transition, $T_{c}$, has been
measured systematically for uniform $Pb$ wires of various widths.  The
effective strength of the disorder characterized by $R_{\Box}$ has
been controlled by the wire thickness $d$.  Before going to a detailed
comparison of the theory with the experiment a few remarks are in
order. The theory described above deals with the universal mechanism
related to large scale distances that are of the order of the thermal
length $L_T\propto \sqrt{D/T}$).  However, a number of other effects
may also influence $T_c$ when the thickness $d$ is decreased.  For
example, the electron states quantization and the interaction of the
electrons with the film's substrate can alter the parameters of the
electron liquid.  These nonuniversal effects of a short range origin
are not addressed by the present theory.  In some systems, e.g., $Mo
Ge$ (see Ref.~\cite{SC2D:Finkelstein8788,SC2D:Graybeal84}), the
discussed effect, originated from the interplay of the Coulomb
interaction and disorder, is dominant, and the theoretical curve
matches the experimental data at $2D$.  Unfortunately, as it is shown
in Fig.~\ref{fg:et2D} the theoretical curve for $Pb$ films does not
follow the experiment. This fact indicates that the effects of a short
range physics are not negligible here.
 To minimize the role of the nonuniversal effects, and make the
comparison between the experiment and theory possible, we proceed in
the following way.  For each width the theoretical curve has been
multiplied by the function $x\left( R_{\Box }\right)
=T_{c}^{2D}(R_{\Box})_{ex}/T_{c}^{2D}(R_{\Box})_{th}$.  This function
is the ratio between the two curves presented in Fig.  \ref{fg:et2D}.
Here, the basic idea is that, because the widths of the wires are
considerably larger than any microscopical scale, the influence of the
short range effects on $T_{c}$ in wires remains the same as in $2D$
films.  In this way, we believe, the effect of the long range physics
determining the crossover from $2D$ to $1D$ systems can be captured by
the present theory. To continue further we have to discuss another
complication. Unlike the case of $2D$ films, the limit of $R_\Box
\rightarrow 0$ for a series of wires with a fixed width is somewhat
ambiguous. For the discussed data, the extrapolation of $T_{c}$ to the
limit $R_{\Box} \rightarrow 0$ at a fixed width yields values that are
not equal to the transition temperature in the bulk limit. (Moreover,
the extrapolated values behave in an irregular way as a function of
the wire width.) Under this circumstance, we have normalized the
theoretical curves in such a way that in the limit $R_{\Box}
\rightarrow 0$ the fitting curves for each width, $a$, start from the
extrapolated $T_{c0}(a)=T_{c}(R_{\Box } \rightarrow 0)$.  After this
normalization procedure and rescaling the theoretical curves by
$x\left( R_{\Box }\right)$, the data for wires of different widths has
been plotted together with the theoretical curves in
Fig.~\ref{fg:et250}.  The fitting parameter $\gamma =-0.16$,
determined from the initial slope of the $T_{c}(R_{\Box})$ in $2D$
films, was the same for all wire widths.  Notice that at $R_\Box
\gtrsim 2000 \Omega$ the suppression of $T_c$ for the wire of the
smallest width is about 1.5 times stronger than for the widest
one. The agreement between theory, i.e.  Eq.~(\ref{eq:Tc}), and
experimental data for all wires of different widths turned out to be
very good.
  To summarize, we have developed a theory that describes the
suppression of the mean-field temperature of the superconducting
transition in amorphous systems. The theory is based on the
consideration of the suppression of the contact attraction due to
phonons, by the dynamically enhanced Coulomb repulsion. It is suitable
for the description of the crossover region between $2D$ and $1D$. By
treating the problem as an eigenvalue problem, we overcame the
difficulties occurring because of the coexistence of different
singularities in the equation determining $T_{c}$.  In order to
compare the available experimental results with the theory, we
analyzed the data in a way that minimizes the role of nonuniversal
effects of a short range origin.  We believe that the theory could
be tested further with superconducting wires fabricated from other
materials, where the initial slope of $T_c (R_\Box)$ is larger than in
$Pb$ films.

It is our pleasure to acknowledge discussions with I.~L.~Aleiner,
V.~Ambegaokar, M.~E.~Gershenson, L.~S.~Levitov, M.~Yu.~Reizer, and
R.~A.~Smith. The research is supported by the DIP Cooperation, the
Israel Science Foundation--Centers of Excellence (MOKKED), and by the
NSF grant DMR-94-16910. YO is grateful for the support by the
Rothschild Fund.

\end{document}